\documentclass[12pt,a4paper,twoside]{article}

\usepackage{amsmath}
\usepackage{amssymb}

\numberwithin{equation}{section}
\newtheorem{theorem}{Theorem}[section]
\newtheorem{lemma}[theorem]{Lemma}
\newtheorem{definition}[theorem]{Definition}
\newtheorem{remark}[theorem]{Remark}
\newtheorem{example}[theorem]{Example}
\newtheorem{proposition}[theorem]{Proposition}

\newenvironment{proof}{\paragraph{Proof.}}{\hfill $\square$\\}
\newenvironment{proof*}{\paragraph{Proof.}}{}

\newcommand{\hk}{\hslash}
\newcommand{\arrow}{\rightarrow}
\newcommand{\map}{\mapsto}
\newcommand{\bb}[1]{\mathbb{#1}}
\newcommand{\diff}[2]{\frac{d#1}{d#2}}
\newcommand{\Diff}[3]{\left . \frac{d#1}{d#2}\right |_{#3}}
\newcommand{\alg}{\mathfrak{g}}
\newcommand{\Alg}{\mathcal{A}}
\newcommand{\C}{\mathcal{C}}
\newcommand{\X}{\mathcal{X}}

\newcommand{\Cm}{\mathbb{C}}
\newcommand{\Z}{\mathbb{Z}}
\newcommand{\si}{\sigma}

\newcommand{\Rm}{\mathbb{R}}
\newcommand{\Km}{\mathbb{K}}
\newcommand{\T}{{\rm T}}
\newcommand{\Time}{\mathbb{T}}
\newcommand{\G}{\mathbb{G}}
\newcommand{\F}{\mathcal{F}}
\newcommand{\V}{\mathcal{V}}
\newcommand{\U}{\mathcal{U}}
\newcommand{\pr}{\partial}
\newcommand{\me}{\geqslant}
\newcommand{\les}{\leqslant}
\newcommand{\bra}[1]{\left (#1\right )}
\newcommand{\brac}[1]{\left [#1\right ]}
\newcommand{\pobr}[1]{\left \{#1\right \}}
\newcommand{\dual}[1]{\left \langle #1 \right \rangle}
\newcommand{\var}[2]{\frac{\delta #1}{\delta #2}}
\newcommand{\pd}[2]{\frac{\partial #1}{\partial #2}}
\newcommand{\ad}{{\rm ad}}
\newcommand{\eqreff}[2]{(\ref{#1}-\ref{#2})}
\newcommand{\res}{{\rm res}}
\newcommand{\e}{\mathcal{E}}

\newcommand{\uf}{{\bf u}}
\newcommand{\D}{\Delta}
\newcommand{\K}{{\bf K}}
\newcommand{\dx}{\,d_{\hk}x}

\newcommand{\tr}{{\rm Tr}}
\newcommand{\alga}{\mathfrak{a}}
\newcommand{\id}{{\rm id}}

\begin{document}

\title{Integrable discrete systems on $\Rm$\\ and related dispersionless systems}

\author{Maciej B\l aszak$^{1}$, Metin G{\" u}rses$^{2}$, Burcu
Silindir$^{2}$\\
and B\l a\.zej M. Szablikowski$^{1}$\\[3mm]
\small (1) Institute of Physics, A. Mickiewicz University\\
\small Umultowska 85, 61-614 Pozna\'n, Poland\\
\small E-mails: blaszakm@amu.edu.pl and bszablik@amu.edu.pl\\[3mm]
\small (2) Department of Mathematics, Faculty of Sciences\\
\small Bilkent University, 06800 Ankara, Turkey\\
\small E-mails: gurses@fen.bilkent.edu.tr and
silindir@@fen.bilkent.edu.tr}
\date{}
\maketitle

\begin{abstract}
The general framework for integrable discrete systems on $\Rm$ in
particular containing lattice soliton systems and their
$q$-deformed analogues is presented. The concept of regular grain
structures on $\Rm$, generated by discrete one-parameter groups of
diffeomorphisms, through which one can define algebras of shift
operators is introduced. Two integrable hierarchies of discrete
chains together with bi-Hamiltonian structures are constructed.
Their continuous limit and the inverse problem based on the
deformation quantization scheme are considered.
\end{abstract}

\section{Introduction}

In recent years of wide interest have become the so-called
integrable $q$-analogues of KP and Toda types hierarchies together
with related Hamiltonian structures, $W$-algebras,
$\tau$-functions, etc. (see \cite{k}-\cite{dmh} and references
therein). The $q$-deformed KP hierarchy ($q$-KP) with the
reductions of $q$-KdV soliton type systems are obtained by means
of pseudo-differential operators defined through $q$-derivative
$\pr_q$ instead of the usual derivative $\pr$ used for ordinary KP
and KdV hierarchies:
\begin{equation*}
\pr u(x) = \pd{u(x)}{x}\qquad\arrow\qquad\pr_qu(x) =
\frac{u(qx)-u(x)}{(q-1)x}.
\end{equation*}
Analogously $q$-deformed Toda hierarchies can be constructed by
means of $q$-shift operators:
\begin{equation*}
E u(x) = u(x+1)\qquad\arrow\qquad E_q u(x)=u(qx).
\end{equation*}
The scheme of the construction of integrable $q$-deformed systems
is based on the classical $R$-matrix formalism that proved very
fruitful for systematic construction of field and lattice soliton
systems \cite{sts}-\cite{bs} as well as dispersionless integrable
field systems \cite{tt}-\cite{sb}. Moreover, the $R$-matrix
approach allows a construction of Hamiltonian structures and
conserved quantities. By integrable systems we understand these
which have infinite hierarchy of symmetries and conserved
quantities.

Having all the above classes of integrable systems, with parallel
schemes of construction, it is interesting how to embed them into
a more general unifying framework. One way of doing this is a
construction of integrable systems on time scales \cite{ah,ggs}. A
time scale $\Time$ is an arbitrary nonempty closed subset of real
numbers. It was introduced to unify all possible intervals on the
real line $\Rm$, like continuous (whole) $\Rm$, discrete $\Z$, and
$q$-discrete $\mathbb{K}_q$ intervals. On a given time scale it is
possible to construct $\D$-derivative (being simultaneously
generalization of the ordinary derivative and the $q$-derivative)
through forward $\si(x)$ and backward $\rho(x)$ jump operators,
where $x\in \Time$ (for all precise definitions see
\cite{ah,ggs}). Assuming the regularity property of $\Time$,
implying that $\rho(\si(x))=x$, one can define an algebra of
Laurent series of $\D$-operators
\begin{equation*}
  \D u(x) = \frac{u(\si(x))-u(x)}{\mu(x)}\qquad \mu(x)\equiv
  \si(x)-x,\qquad x\in \Time
\end{equation*}
or shift operators $Eu(x)=u(\si(x))$, leading to the construction
of integrable systems on time scales. Defining suitable inner
products in this algebras, one can construct additionally
conservation laws. In such a formulation, dynamical fields
$u:\Time\arrow \Rm$ are mappings from a time scale to real
numbers.

The main goal of this work is the formulation of a general
unifying framework of integrable discrete systems, in such a way
that the domain of dynamical fields $u$ is always $\Rm$. We also
consider the continuous limit and the inverse procedure. In the
second section we introduce the concept of regular grain structure
on $\Rm$ defined by discrete one-parameter groups of
diffeomorphisms $\si_{m\hk}(x)$. Then, the shift operator can be
constructed through formal jump operator $\si(x)=\si_{\hk}(x)$. In
this section elements of geometric scheme are defined as
appropriate functionals, duality maps, adjoint operators, etc. A
class of discrete systems is chosen in such a way that the limit
$\hk\arrow 0$ is dispersionless. In the third section, using the
formalism of classical $R$-matrices, we construct two integrable
hierarchies of discrete chains being counterparts of the original
infinite-field Toda and modified Toda chains. Additionally
bi-Hamiltonian structures are constructed. In the next section the
concept of the continuous limit, which in our case becomes the
dispersionless limit, is explained. Further, in the fifth section,
the theory of dispersionless chains, being dispersionless limits
of discrete chains together with bi-Hamiltonian structures is
presented. In the sixth section the inverse problem to the
dispersionless limit is considered. It is based on the scheme of
deformation quantization formalism introduced in \cite{bs}. As a
result we show that there is a class of gauge equivalent
integrable discrete systems being dispersive counterparts of
dispersionless systems considered earlier. We end the paper with
some final comments.

\section{One-parameter regular grain structures on $\Rm$}

The main aim of this article is the formulation of a general
theory of integrable discrete systems on $\Rm$, that will contain
lattice soliton systems as well as $q$-discrete systems as
particular cases. This theory will be illustrated by integrable
discrete chains being infinite-field systems.

Maps $\si:\Rm\arrow\Rm$ and $\rho:\Rm\arrow\Rm$ will be called the
forward and backward jump operators, respectively. These names are
only a convention as we do not assume that $\si(x)\me x$ and
$\rho(x)\les x$ for arbitrary $x\in\Rm$.  In $n\in\Z_+$ forward
steps the point $x\in\Rm$ is mapped to the point $\si^n(x)$, where
$\si^n$ is the $n$-times composition of forward jump operator
$\si$. Respectively, in $n$ backward steps, $x$ is mapped to the
point $\rho^n(x)$. Then, the range of possible points to which we
can map $x$ by forward and backward steps (including $x$) is given
by
\begin{equation*}
  \G_x:=
  \pobr{\rho^n(x):n\in\Z_+}\cup\pobr{x}\cup\pobr{\si^n(x):n\in\Z_+}.
\end{equation*}
Hence, to each point $x$ of $\Rm$ a set $\G_x$ is associated. The
union of all $\G_x$ is given by $\G:=\bigcup_{x\in\Rm}\G_x$.

\begin{definition}
We will say that $\G$ defines the grain structure on $\Rm$. We
will call it the regular grain structure, if there exist inverse
maps $\si^{-1}$ and $\rho^{-1}$, such that $\si(x)=\rho^{-1}(x)$
and $\rho(x)=\si^{-1}(x)$ for all $x\in\Rm$.
\end{definition}

So, to define the regular grain structure on $\Rm$ one needs only
one forward jump operator $\si$ being bijection, as the backward
operator is given by $\si^{-1}$. Then,
$\G_x=\pobr{\si^{n}(x):n\in\Z}$, where we assumed that
$\si^0\equiv\id_\Rm$. Besides, bijective $\si$ defines a discrete
one-parameter group of bijections on $\Rm$: $\Z\ni m\map
\pobr{\si_m:\Rm\arrow\Rm}$, such that $\si_m:=\si^m$, and vice
versa each one-parameter group of bijections on $\Rm$ defines the
regular grain structure on $\Rm$ with the forward jump operator
defined by $\si:=\si_1$. Notice that the regular grain structure
introduces equivalence classes between points of $\Rm$, such that
$x\sim y$ if $\G_x=\G_y$ ($x,y\in\Rm$), i.e. there exists $k\in\Z$
such that $y=\si^k(x)$.

Further on, we introduce a regular grain structure $\G$ on $\Rm$
by one-parameter group of diffeomorphisms instead of bijections,
which is necessary as we will deal with differential geometry of
infinite-dimensional systems with smooth dynamical fields. Let $\Z
\ni m\map \si_{m\hk}$ be a discrete one-parameter group of
diffeomorphisms on $\Rm$: $\si_{m\hk}:\Rm\arrow\Rm$, i.e
\begin{align*}
    \si_0(x)=x\quad \text{and}\quad
    \si_{m\hk}\bra{\si_{n\hk}(x)}=\si_{(m+n)\hk}(x)\qquad m,n\in \Z,
\end{align*}
where $\hk>0$ is some deformation parameter. It follows that
$(\si_{n\hk})^{-1}(x)=\si_{-n\hk}(x)$. The continuous
one-parameter group of diffeomorphisms ($\Rm\ni t\map \si_t$) can
be completely determined by its infinitesimal generator
$\X(x)\pr_x$ being a vector field on $\Rm$. We assume that the
component $\X(x)$ is smooth on $\Rm$ except at most at a finite
number of points. Then,
\begin{equation}\label{fl}
    \X(x) = \Diff{\si_t(x)}{t}{t=0} \qquad\Leftrightarrow\qquad
    \diff{\si_t(x)}{t} = \X\bra{\si_t(x)},
\end{equation}
where $t\in\Rm$. Arbitrary $\X\pr_x$ generates a continuous
one-parameter group of diffeomorphisms only when it is a {\it
complete} vector field, for which maximal integrals are defined on
the whole $\Rm$, i.e. $\Rm$ is a domain of the mapping $t\map
\si_t$. In such a case the above discrete one-parameter group is
well defined as it is enough to consider subgroup $\Z$ of $\Rm$.
Incomplete $\X\pr_x$ might still well define a discrete group of
diffeomorphisms, if $\hk$ is properly chosen.

\begin{lemma}\label{lem}
Let $\si_t(x)$ be a one-parameter group of diffeomorphisms
generated by $\X(x)\pr_x$. Then, the following relation is valid
\begin{equation}\label{rel}
\X(x)\diff{\si_t(x)}{x} = \X\bra{\si_t(x)}.
\end{equation}
\end{lemma}
\begin{proof}
>From \eqref{fl} one observes that
$\X\bra{\si_{s+t}(x)}=\diff{\si_{s+t}(x)}{s}$. Hence, the
following relation is valid
\begin{equation*}
\X\bra{\si_s(x)}\diff{\si_{s+t}(x)}{\si_s(x)} =
\X\bra{\si_{s+t}(x)}.
\end{equation*}
Since it can be obtained from \eqref{rel} by acting on its both
sides with $\si_s$, the proposition is completed.
\end{proof}

Now, we establish a phase space related to discrete systems under
considerations. Let
\begin{equation*}
\uf:=(u_0(x),u_1(x),u_2(x),...)^\T
\end{equation*}
be a infinite-tuple of smooth functions $u_i:\Rm\arrow \Km$, $x\map
u_i(x)$ with values in $\Km=\Rm$ or $\Cm$. Additionally we assume
that $u_i$ depend on an appropriate set of evolution parameters, and
so $u_i$ are dynamical fields. Let $\U$ be a linear topological
space, with local independent coordinates $\uf\bra{\si_{m\hk}(x)}$
for all $m\in\Z$, defining our infinite-dimensional phase space. We
will use the following notation
\begin{equation*}
  (E^mu)(x)\equiv u\bra{\si_{m\hk}(x)}.
\end{equation*}
Let $\C$ be the algebra over $\Km$ of functions on $\U$ of the
form
\begin{equation}\label{pol}
  f\brac{\uf} := \sum_{m\me 0}\sum_{i_1,\ldots,i_m\me 0}
  \sum_{s_1,\ldots,s_m\in\Z}
  a^{i_1i_2\ldots i_m}_{s_1s_2\ldots s_m}
(E^{s_1}u_{i_1})(E^{s_2}u_{i_2})\cdots (E^{s_m}u_{i_m})
\end{equation}
that are polynomials in $\uf\bra{\si_{m\hk}(x)}$ of finite order,
with coefficients $a^{i_1i_2\ldots i_m}_{s_1s_2\ldots s_m}\in\Km$.
This algebra can be extended into operator algebra
$\C\brac{E,E^{-1}}$ ($\C[x,y,...]$ stands for the linear space of
polynomials in $x,y,...$ with coefficients from $\C$), where $E$
is a shift operator compatible with the grain structure defined by
$\si_\hk(x)$, i.e.
\begin{equation*}
  E^mu(x):=(E^mu)(x)=u\bra{\si_{m\hk}(x)}\qquad m\in\Z,
\end{equation*}
where $u(x)$ is some field. As $\si_\hk(x)$ is an element of
one-parameter group of diffeomorphisms, hence
\begin{equation}\label{fact1}
  \si_\hk(x)= e^{\hk\X(x)\pr_x}x\qquad\Leftrightarrow\qquad
  e^{\hk\X(x)\pr_x}u(x)=u\bra{e^{\hk\X(x)\pr_x}x}
\end{equation}
is valid if $\X(x)\pr_x$ is a complete vector field and $u(x)$ is
a smooth function. Thus, the shift operator $E$ can be identified
with $e^{\hk\X(x)\pr_x}$, i.e.
\begin{equation}\label{fact}
E^m\equiv e^{m\hk\X(x)\pr_x}.
\end{equation}

\begin{example}\label{example}
Consider vector fields on $\Rm$ of the form
$\X(x)\pr_x=x^{1-n}\pr_x$ for $n\in \Z$. For $n=0$ integrating
\eqref{fl} one finds that
\begin{equation*}
  \si_t(x) = e^tx \qquad\Rightarrow\qquad \si_{m\hk}(x) =
  e^{m\hk}x = q^mx\qquad q\equiv e^\hk,
\end{equation*}
which is defined for all $t\in\Rm$ and so $\X\pr_x=x\pr_x$ is a
complete vector field. When $n=0$, we will deal with systems of
'$q$-discrete' type. When $n\neq 0$, in general, $\si_t(x)$ can be
find only in the implicit form
\begin{equation*}
  \bra{\si_t(x)}^n = x^n + n t.
\end{equation*}
For $n=1$ we have
\begin{equation*}
  \si_t(x) = x+t \qquad\Rightarrow\qquad \si_{m\hk}(x) =
  x+m\hk,
\end{equation*}
and $\X\pr_x=\pr_x$ is obviously complete. In this case we will
deal with systems of 'lattice' type. For $n=-1$ the related vector
field $\X\pr_x=x^2\pr_x$ is incomplete as $t\neq \frac{1}{x}$:
\begin{equation*}
  \si_t(x) = \frac{x}{1 - t x}\qquad\Rightarrow\qquad
  \si_{m\hk}(x) = \frac{x}{1 - m\hk\, x}.
\end{equation*}
However, if $x\neq\frac{1}{m \hk}$, the related discrete
one-parameter group of diffeomorphisms is well defined. When $n$
is odd, we can always define a discrete one-parameter group of
diffeomorphisms generated by $\X\pr_x=x^{1-n}\pr_x$. Another
example of incomplete vector field is given for $n=2$
($\X\pr_x=\frac{1}{x}\pr_x$) since we have
\begin{equation*}
  \si_t(x) = {\rm sign}(x)\sqrt{x^2+2t}
\end{equation*}
with restriction $t\me -\frac{1}{2}x^2$. The vector fields
$\X\pr_x = x^{1-n}\pr_x$ of this kind should be excluded from
further considerations or properly extended over complex plane
$\Cm$.
\end{example}

A space $\F=\pobr{F:\U\arrow\Km}$ of functions on $\U$ is defined
through linear functionals
\begin{align}\label{funct}
\int(\cdot)\dx:\C\arrow\Km\qquad f[\uf]\map F(\uf):=\int
f[\uf]\dx,
\end{align}
such that
\begin{equation}\label{rest}
    \int Ef[\uf]\dx = \int f[\uf]\dx,
\end{equation}
where $\int\dx$ is a formal integration symbol. The restriction
\eqref{rest} entails the form of adjoint with respect to the
duality map which will be defined in a moment.

\begin{definition}\label{def}
The explicit form of appropriate functionals can be introduced in
two ways.
\begin{itemize}
  \item[(i)] A discrete representation is defined as
\begin{equation}\label{fun1}
  F(\uf) = \int f[\uf]\dx := \hk
  \sum_{n\in\Z}f\brac{\uf\bra{\si_{n\hk}(x)}}.
\end{equation}
  \item[(ii)] A continuous representation is given as
\begin{equation}\label{fun2}
  F(\uf) = \int f[\uf]\dx :=
  \int_{-\infty}^{\infty}f[\uf(x)]\frac{dx}{\X(x)},
\end{equation}
where we assume that $u_i(x)$ vanishes as $|x|\arrow\infty$ (if
$\X(x)\arrow 0$ for $|x|\arrow\infty$, then $u_i(x)$ must vanish
faster then $\X(x)$ does). The above integral in general is
improper, so additionally we have to assume that $u_i(x)$ behave
properly as $x$ tends to critical points $x_c$ of $\X(x)$
($\X(x_c)=0$). Then, evaluating the integral we take its principal
value.
\end{itemize}
 \end{definition}

When it is not necessary to differentiate between the above
representations, we will use only the formal integration symbol
$\int\dx$. We have explicitly defined the functionals in two ways
reflecting two different approaches developed for the lattice
soliton systems. The first one is with the domain of dynamical
fields $\Z$ \cite{bm,o}, the second one with $\Rm$ \cite{bs,cdz}.
So, the functionals \eqref{fun1} and \eqref{fun2} are appropriate
generalizations of these two approaches.

\begin{proposition} Both functionals from Definition
\ref{def} are well defined and satisfy \eqref{rest}.
\end{proposition}
\begin{proof}Both functionals are trivially linear. The discrete one
satisfies \eqref{rest} since we can freely change the boundaries
because we sum over the whole $\Z$. For the second functional we
have that
\begin{align*}
\int Ef[\uf]\dx &=
\int_{-\infty}^{\infty}f[\uf\bra{\si_\hk(x)}]\frac{dx}{\X(x)}
=\int_{-\infty}^{\infty}f[\uf(x)]\diff{\si_{-\hk}(x)}{x}\frac{dx}{\X\bra{\si_{-\hk}(x)}}\\
&=\int_{-\infty}^{\infty}f[\uf(x)]\frac{dx}{\X(x)} =\int
f[\uf]\dx,
\end{align*}
where one obtains the second equality by the change of variables
$x\map \si_\hk(x)$, while the next one follows from Lemma
\ref{lem}.
\end{proof}

A vector field on $\U$ is given by a system of
differential-difference equations, the difference one with respect
to the grain structure defined by $\si_\hk$ and the first order
differential one with respect to the evolution parameter $t$,
\begin{equation}\label{flow}
\uf_t = \K(\uf),
\end{equation}
where $u_t:=\pd{u}{t}$ and $\K(\uf) := \frac{1}{\hk}(K_1[\uf],
K_2[\uf], ...)^\T$ with $K_i[\uf]\in\C$. The class of the discrete
systems is chosen in such the way that in the continuous limit
$\hk\arrow 0$ we obtain systems of hydrodynamic type (see Section
4). This assumption explains the appearance of the factor
containing $\hk$ in $\K$.

Let $\V$ be a linear space over $\Km$, of all such vector fields
on $\U$. Then the dual space $\V^*$ is a space of all linear maps
$\eta:\V\arrow\Km$. The action of $\eta\in \V^*$ on $\K\in \V$ can
be defined through a duality map (bilinear functional)
$\dual{\cdot,\cdot}: \V^*\times \V \arrow \Km$ given by functional
\eqref{funct} as
\begin{equation}\label{dual}
 \dual{\eta, \K} = \int \sum_{i=0}^\infty \eta_i K_i\dx
  = \int \eta^\T\cdot \K\dx,
\end{equation}
where components of $\eta:= (\eta_1, \eta_2, ...)^\T$ belong to
$\C$. With respect to the duality map \eqref{dual} one finds that
the adjoint of $E^m$ is equal to $E^{-m}$, i.e.
$\bra{E^m}^\dag=E^{-m}$.

\begin{proposition}
The differential
\begin{equation*}
dF(\uf) = \bra{\var{F}{u_0}, \var{F}{u_1}, ...}^\T\in\V^*
\end{equation*}
of a functional $F(\uf)=\int f[\uf]\dx$, such that its pairing
with $\K\in\V$ assumes the usual Euclidean form
\begin{equation}\label{eucl}
F'[\K]= \dual{dF,\K} = \int \sum_{i=0}^\infty \var{F}{u_i} (u_i)_t\dx,
\end{equation}
where $F'[\K]$ is the directional derivative, is defined by
variational derivatives of the form
\begin{equation*}
\var{F}{u_i} := \sum_{m\in
\bb{Z}}E^{-m}\pd{f[\uf]}{u_i\bra{\si_{m\hk}(x)}}.
\end{equation*}
\end{proposition}
\begin{proof}Let $\uf_t = \K(\uf)$, then
\begin{equation*}
  F'(\uf)[\uf_t] \equiv \diff{F(\uf)}{t} = \int
  \sum_{i=0}^\infty\sum_{m\in\Z}
  \pd{f[\uf]}{u_i\bra{\si_{m\hk}(x)}}\diff{u_i\bra{\si_{m\hk}(x)}}{t}\dx
  = \int\sum_{i=0}^\infty \var{F}{u_i} (u_i)_t\dx,
\end{equation*}
where the last equality follows from \eqref{rest}.
\end{proof}

Further we will be interested in bi-vector fields on $\U$ defined
through linear operators $\pi:\V^*\arrow\V$, which in a local
representation are matrices with coefficients from $\C[E,E^{-1}]$
multiplied by $\frac{1}{\hk}$. An operator $\pi$ is a Poisson
operator (tensor) if the bilinear bracket
\begin{equation*}
  \pobr{H, F}_\pi = \dual{dF, \pi dH}\qquad F, H\in
  \F
\end{equation*}
is a Poisson bracket.

\begin{remark}
It is important to mention that the particular choice of the
algebra $\C$, and consequently the algebra $\C[E,E^{-1}]$,
determines the class of discrete systems considered, which in the
limit $\hk\arrow 0$ tend to differential systems of first order,
i.e. dispersionless ones. Alternative approach to the construction
of discrete systems on $\Rm$ with the grain structure $\G$ is
based on the use of $\D$-derivative, instead of the shift
operation, given by
\begin{equation*}
  \D u(x) := \frac{(E-1)u(x)}{(E-1)x} =
  \frac{u\bra{\si_\hk(x)}-u(x)}{\mu_\hk(x)}\qquad \mu_\hk(x)\equiv
    \si_\hk(x)-x.
\end{equation*}
In this case, the algebra $\C$ is composed of polynomials in
$\D^mu$ ($m=0,1,...$) and the operator algebra is given by
$\C[\D]$. Consequently the restriction \eqref{rest} on the
functional is replaced by
\begin{equation}\label{rest1}
  \int'\D f[u]\dx =0,
\end{equation}
which entails that $\D^\dag = \D E^{-1}$ with respect to the
duality map generated by this functional. Prime in $\int'$ is used
to differ the functional satisfying property \eqref{rest1} from the
functional satisfying property \eqref{rest}. Nevertheless, both functionals
are interrelated by the relation
\begin{equation*}
  \int'(\cdot)\dx = \int(\cdot)\mu_\hk(x)\dx,
\end{equation*}
which is a consequence of the restrictions imposed on them.
Contrary to the previous case the continuous limit of discrete
systems from the alternative approach with $\D$-operation gives
dynamical field systems with dispersion and is not considered in
this article.
\end{remark}

\section{$R$-matrix approach to integrable discrete systems
on~$\Rm$}\label{sect}

Now, we are ready for the construction of integrable discrete
systems following from the scheme of classical $R$-matrix
formalism parallel to the one used in the case of lattice soliton
systems \cite{bm,b,bs}.

On $\Rm$ with the grain structure $\G$ defined by some
diffeomorphism $\si_\hk$ we introduce the algebra of shift
operators with finite highest order:
\begin{equation}\label{alg}
  \alg = \alg_{\me k-1} \oplus \alg_{< k-1} = \pobr{\sum^N_{i\me k-1}
u_i(x)\e^i} \oplus \pobr{\sum_{i< k-1} u_i(x)\e^i}
\end{equation}
where
\begin{equation}\label{mult}
\e^m u(x) = (E^m u)(x)\e^m\equiv
u(\si_{m\hk}(x))\e^m\qquad\si_{m\hk}:=\si_\hk^m\qquad m\in \Z
\end{equation}
and $u_i(x)$ are smooth dynamical fields.

\begin{proposition}
The multiplication operation in $\alg$ defined by \eqref{mult} is
non-commutative and associative.
\end{proposition}
\begin{proof}
Non-commutativity is obvious. Associativity follows from
straightforward calculation and from the fact that $\si_{m\hk}$ is
a one-parameter group of diffeomorphisms.
\end{proof}

 The Lie structure on $\alg$ is introduced through the commutator
\begin{equation*}
  [A,B] = \frac{1}{\hk}\bra{AB - BA}\qquad A,B\in\alg.
\end{equation*}
The subsets $\alg_{\me k-1}$ and $\alg_{< k-1}$ of $\alg$ are Lie
subalgebras only for $k=1$ and $k=2$. As a result, we can define
the classical $R$-matrices $R = P_{\me k-1} - \frac{1}{2}$,
through appropriate projections, and related Lax hierarchies:
\begin{equation}\label{laxh}
L_{t_n} = \brac{\bra{L^n}_{\me k-1},L}= \pi_0 dH_n =
\pi_1 dH_{n-1}\qquad n\in \Z_+\qquad k=1,2,
\end{equation}
of infinitely many mutually commuting systems. The evolution
equations from \eqref{laxh} are generated by powers of appropriate
Lax operators $L\in \alg$ of the form:
\begin{align}
\label{l1} k=1:\qquad L &= \e+u_0+u_1\e^{-1}+u_2\e^{-2}+...=
\e+\sum_{i\me 0}u_{i}\e^{-i}\\
\label{l2} k=2:\qquad L &= u_0\e+u_1+u_2\e^{-1}+u_3\e^{-2}+...=
\sum_{i\me 0}u_{i}\e^{1-i}.
\end{align}

Then, the first chains from \eqref{laxh} are:
\begin{align}
\label{toda} (u_i)_{t_1} &= \frac{1}{\hk}\brac{(E-1)u_{i+1}+u_i(1-E^{-i})u_0}\\
\notag (u_i)_{t_2} &= \frac{1}{\hk}\left[(E^2-1)u_{i+2}+Eu_{i+1}(E+1)u_0-u_{i+1}(E^{-i}+E^{-i-1})u_0\right.\\
\notag         &\qquad\ \left.+u_i(1-E^{-i})u_0^2+u_i(E+1)(1-E^{-i})u_1\right]\\
\notag          & \vdots
\end{align}
for $k=1$, and
\begin{align*}
(u_i)_{t_1} &= \frac{1}{\hk}\brac{u_0Eu_{i+1}-u_{i+1}E^{-i}u_0}\\
(u_i)_{t_2} &= \frac{1}{\hk}\left[u_0Eu_0E^2u_{i+2}-u_{i+2}E^{-i-1}u_0E^{-i}u_0\right.\\
        &\qquad\ \left.
        +u_0(E+1)u_1Eu_{i+1}-u_{i+1}E^{-i}u_0(E^{1-i}+E^{-i})u_1\right]\\
& \vdots
\end{align*}
for $k=2$. Here and further on the shift operators $E^m$ in
evolution equations and conserved quantities act only on the
nearest field on the right and in Poisson operators act on
everything on the right of the symbol $E^m$ inside and outside the
operator.

\begin{example}The lattice case: $\X=1$.

Let $\hk=1$. The first chains of evolution equations from \eqref{laxh} have
the form:
\begin{align*}
k=1:\qquad u_i(x)_{t_1} &= u_{i+1}(x+1)-u_{i+1}(x) + u_i(x)\bra{u_0(x)-u_0(x-i)}\\
k=2:\qquad u_i(x)_{t_1} &= u_0(x)u_{i+1}(x+1)-u_0(x-i)u_{i+1}(x).
\end{align*}
These are Toda and modified Toda chains, respectively.
\end{example}

\begin{example}The $q$-discrete case: $\X=x$ ($q\equiv e^\hk$).

In this case the same evolution equations are
\begin{align*}
k=1:\qquad u_i(x)_{t_1} &= u_{i+1}(q x)-u_{i+1}(x) + u_i(x)\bra{u_0(x)-u_0(q^{-i} x)}\\
k=2:\qquad u_i(x)_{t_1} &= u_0(x)u_{i+1}(q x) -
                u_0(q^{-i} x) u_{i+1}(x),
\end{align*}
where the constant factor $\hk$ is absorbed into the evolution parameter
$t_1$ through simple rescaling. These are $q$-deformed analogues
of the chains from the previous example.
\end{example}

In this work we do not consider finite-field reductions of
\eqref{laxh} as the procedure is straightforward following
\cite{bm,bs}.

To construct Hamiltonian structures for \eqref{laxh} at first one
has to define an appropriate inner product on $\alg$.
\begin{definition}
Let $\tr:\alg\arrow\Km$ be a trace form, being a linear map, such
that
\begin{equation*}
  \tr(A) :=  \int \res (A\e^{-1})\dx,
\end{equation*}
where $\res (A\e^{-1}):=a_0$ for $A=\sum_i a_i\e^i$. Then, the
bilinear map $(\cdot,\cdot):\alg\times\alg\arrow\Km$ defined as
\begin{equation}\label{dm}
\bra{A,B}:= \tr\bra{A B}
\end{equation}
is an inner product on $\alg$.
\end{definition}
\begin{proposition}
The inner product \eqref{dm} is nondegenerate, symmetric and
$\ad$-invariant, i.e.
\begin{equation*}
  \bra{\brac{A,B},C} = \bra{A,\brac{B,C}}\qquad A,B,C\in\alg.
\end{equation*}
\end{proposition}
\begin{proof}
The nondegeneracy of \eqref{dm} is obvious. The symmetricity
follows from \eqref{rest}. The $\ad$-invariance is a consequence
of the associativity of multiplication operation in $\alg$.
\end{proof}

Next, the differentials $dH(L)$ of functionals $H(L)\in \F(\alg)$
for \eqreff{l1}{l2} have the form:
\begin{align*}
  k=1:\qquad dH&= \sum_{i\me 0}\e^i\var{H}{u_i}\\
  k=2:\qquad dH&= \sum_{i\me 0}\e^{i-1}\var{H}{u_i},
\end{align*}
which follows from the assumption that inner product on $\alg$ is
compatible with \eqref{eucl}, i.e.
\begin{equation*}
  \bra{dH,L_t} = \int\sum_{i=0}^\infty \var{H}{u_i} (u_i)_t\dx.
\end{equation*}
Then, the bi-Hamiltonian structure of the Lax hierarchies
\eqref{laxh} is defined through the compatible (for fixed $k$)
Poisson tensors given by
\begin{align*}
k=1,2:\qquad\pi_0: dH\map
\brac{L,(dH)_{<k-1}}+\bra{\brac{dH,L}}_{< 2-k}
\end{align*}
and
\begin{align*}
k=1:\qquad\pi_1: dH&\map \frac{1}{2}\bra{\brac{L,\bra{L dH + dH
L}_{<0}}
+ L\bra{\brac{dH,L}}_{<1}+\bra{\brac{dH,L}}_{<1}L}\\
&\qquad +\hk\brac{(E+1)(E-1)^{-1}\res\bra{\brac{dH,L}\e^{-1}},L}\\
k=2:\qquad\pi_1: dH&\map \frac{1}{2}\bra{\brac{L,\bra{L dH + dH
L}_{<1}} + L\bra{\brac{dH,L}}_{<0}+\bra{\brac{dH,L}}_{<0}L},
\end{align*}
where the operation $(E-1)^{-1}$ must be understood formally as
the inverse of $(E-1)$ and one can show that
$(E+1)(E-1)^{-1}=\sum_{i=1}^\infty(E^{-i}-E^i)$. The appropriate
Hamiltonians, being conserved quantities, are
\begin{equation*}
  H_n(L) = \frac{1}{n+1}\tr\bra{L^{n+1}}\qquad dH_n(L)=L^n
\end{equation*}
and the explicit bi-Hamiltonian structure of \eqref{laxh} is given
by
\begin{equation*}
  (u_i)_{t_n} = \sum_{j\me 0}\pi_0^{ij}\,\var{H_n}{u_j}
  = \sum_{j\me 0}\pi_1^{ij}\,\var{H_{n-1}}{u_j}\qquad i\me 0.
\end{equation*}
The Poisson tensors for $k=1$ are
\begin{align*}
\pi_0^{ij} &= \frac{1}{\hk}\brac{E^ju_{i+j}-u_{i+j}E^{-i}}\\
\pi_1^{ij} &= \frac{1}{\hk}\Big[
\sum_{k=0}^{i}\bra{u_kE^{j-k}u_{i+j-k}-u_{i+j-k}E^{k-i}u_k +
u_i\bra{E^{j-k}-E^{-k}}u_j}\\
&\qquad + u_i\bra{1-E^{j-i}}u_j +
E^{j+1}u_{i+j+1}-u_{i+j+1}E^{-i-1}\Big]
\end{align*}
together with the hierarchy of Hamiltonians in the form
\begin{align*}
    H_0 &= \int u_0\dx\\
    H_1 &= \int \bra{u_1+\frac{1}{2}u_0^2}\dx\\
    H_2 &= \int
    \bra{u_2+u_0(E+1)u_1+\frac{1}{3}u_0^3}\dx\\
    &\vdots\ .
\end{align*}
For $k=2$ the first Poisson tensor has the following form
\begin{align*}
&\pi_0^{10} = \frac{1}{\hk}(1-E^{-1})u_0\qquad \pi_0^{01} =
\frac{1}{\hk}u_0(E-1)\\
&\pi_0^{ij} = \frac{1}{\hk}\brac{E^{j-1}u_{i+j-1}-u_{i+j-1}E^{1-i}}\quad i,j\me
2,
\end{align*}
with all remaining $\pi_0^{ij}$ equal zero, the second one
is
\begin{align*}
\pi_1^{ij} = \frac{1}{\hk}\Big[\sum_{k=0}^{i-1}
\bra{u_kE^{j-k}u_{i+j-k}-u_{i+j-k}E^{k-i}u_k} +
\frac{1}{2}u_i(E^{1-i}-1)(E^{j-1}+1)u_j\Big]
\end{align*}
and the first Hamiltonians are
\begin{align*}
  H_0 &= \int u_1\dx\\
  H_1 &= \int \bra{\frac{1}{2}u_1^2+u_0Eu_2}\dx\\
  H_2 &= \int \bra{\frac{1}{3}u_1^3+u_0Eu_0E^2u_3+u_0u_1Eu_2+u_0Eu_1Eu_2}\dx\\
  &\vdots\ .
\end{align*}

\section{The continuous limit}

The aim of this section is to consider the limit $\hk\arrow 0$ of discrete systems
\eqref{flow}. The class of this systems is determined by the
choice of the algebra $\C$. Let us assume that the dynamical
fields from $\C$ depend on $\hk$ in such a way that the expansion,
with respect to $\hk$ near zero, is of the form
\begin{equation*}
  u_i(x) = u_i^{(0)}(x) + u_i^{(1)}(x)\hk + O\bra{\hk^2},
\end{equation*}
i.e. $u_i$ in the limit $\hk\arrow 0$ tends to $u_i^{(0)}$. In
further considerations instead of $u_i^{(0)}$ we will still use
$u_i$. In the continuous limit $\C$ becomes the algebra, denoted
by $\C_0$, of polynomial functions in $u_i(x)$:
\begin{equation*}
  \C_0\ni f(\uf) := \sum_{m\me 0}\sum_{i_1,\ldots,i_m\me 0}
   a^{i_1i_2\ldots i_m}\, u_{i_1}(x)u_{i_2}(x)\cdots u_{i_m}(x).
\end{equation*}
In general, the limit of discrete systems \eqref{flow} does not
have to exist. To take the limit, one should first expand the
coefficients of $\K(\uf)$ into a Taylor series with respect to
$\hk$ near $0$, i.e.
\begin{equation*}
  E^mu = e^{m\hk\X\pr_x}u = u+m\hk\X u_x +
  \frac{m^2}{2}\hk^2\bra{\X\X_xu_x+\X^2u_{2x}} + O\bra{\hk^3}.
\end{equation*}
Thus, the continuous limit of \eqref{flow} exists only if in the
above expansion zero order terms in $\hk$ will mutually cancel. In
this case, as $\hk\arrow 0$, the discrete systems \eqref{flow} go
to the systems of hydrodynamic type given in the following form
\begin{equation}\label{dflow}
\uf_t = \X\,{\bf A}(\uf) \uf_x,
\end{equation}
where ${\bf A}(\uf)$ is the matrix with coefficients from $\C_0$,
and the continuous limit is indeed the dispersionless limit.

\begin{proposition}
Assuming that fields $u_i(x)$ vanish as $|x|\arrow\infty$, in the
continuous limit, the functionals from Definition \ref{def} are
given by
\begin{equation}\label{fun}
\int(\cdot)\,d_0x:\C_0\arrow\Km\qquad f[\uf]\map F(\uf)=\int
f(\uf)\,d_0x = \int_{-\infty}^{\infty}f(\uf(x))\frac{dx}{\X(x)}.
\end{equation}
\end{proposition}
\begin{proof}
For the continues case \eqref{fun2} the proof is straightforward.
In the case of discrete functionals \eqref{fun1}, proceeding
analogously as for Riemann integral, we have
\begin{align*}
  \int f[\uf]\,d_0x &\equiv \lim_{\hk\arrow 0} \int f[\uf]\dx =
  \lim_{\hk\arrow 0}
  \sum_{n\in\Z}\hk\,f\brac{\uf\bra{\si_{n\hk}(x)}}\\
  &= \lim_{\hk\arrow 0} \sum_{n\in\Z}
  f\brac{\uf\bra{\si_{n\hk}(x)}} \bra{\frac{\mu_\hk(x)}{\hk}}^{-1}
  \mu_\hk(x)
  = \int_{-\infty}^{\infty}f(\uf(x))\frac{dx}{\X(x)}.
\end{align*}
\end{proof}

Then, bi-vectors $\pi$ are matrices with coefficients of the
operator form $a\X\pr_xb$, where $a,b\in\C_0$. With respect to the
duality map defined by the 'dispersionless' functional \eqref{fun} the
adjoint of the operator $\pr_x$ is given as
\begin{equation}\label{ad1}
  (\pr_x)^\dag = \frac{\X_x}{\X}-\pr_x.
\end{equation}
Consequently, the variational derivatives of functionals $F = \int
f d_0x = \int_{-\infty}^{\infty}f\frac{dx}{X}$ are given by the
derivatives of densities $f$ with respect to the fields $u_i$,
i.e.
\begin{equation*}
\var{F}{u_i} = \pd{f}{u_i}.
\end{equation*}

\begin{example}
The dispersionless limit of the system \eqref{toda} together with its
Hamiltonian structure with respect to the first Poisson tensor is given by
\begin{equation}\label{dtoda}
  (u_i)_{t_1} = \X \brac{(u_{i+1})_x + i u_i (u_0)_x} =
  \pi_0^{ij}\,\var{H_1}{u_j},
\end{equation}
where
\begin{align*}
\pi_0^{ij} = j\X\pr_xu_{i+j}+iu_{i+j}\X\pr_x\qquad\text{and}\qquad
H_1 = \int \bra{u_1+\frac{1}{2}u_0^2}d_0x.
\end{align*}
\end{example}

The Hamiltonian representation of the systems \eqref{dflow} with
the functional \eqref{fun} following directly from the continuous
limit and leads to the nonstandard form with the adjoint operation
for differential operator given by \eqref{ad1}. A more natural
representation is the one with the components $\X(x)$ included in
the densities of functionals given in the standard form
\begin{equation*}
F(\uf) = \int_{-\infty}^{\infty}\X(x)^{-1}f(\uf(x))\,dx\equiv
\int_{-\infty}^{\infty}\varphi(\uf(x))\,dx,
\end{equation*}
for which the variational derivatives preserve the form
$\var{F}{u_i} = \pd{\varphi}{u_i}$. As a consequence, bi-vectors
$\pi$ from the previous representation must be multiplied on the
right-hand side by $\X$. Now, the adjoint of the operator $\pr_x$
takes the standard form $(\pr_x)^\dag = -\pr_x$. Therefore, in
what follows we will use only the natural Hamiltonian
representation of dispersionless systems \eqref{dflow}.

\begin{example}
The natural Hamiltonian structure of
\eqref{dtoda} is given by
\begin{align*}
\pi_0^{ij} = j\X\pr_x\X u_{i+j} + i u_{i+j}\X\pr_x\X
\qquad\text{and}\qquad H_1 = \int_{-\infty}^\infty
\X^{-1}\bra{u_1+\frac{1}{2}u_0^2}dx.
\end{align*}
\end{example}

In the next section we will consider the $R$-matrix formalism of
the dispersionless systems \eqref{dflow}, that can be considered
 as the continuous limit of the
formalism presented in Section 3. That it is really the case
would be clear in Section 6.

\section{$R$-matrix approach to integrable dispersionless systems on~$\Rm$}

The theory of classical $R$-matrices on commutative algebras, with
the multi-Hamiltonian formalism, was given in \cite{li}. Here we
follow the particular scheme of $R$-matrix parallel to the one
developed in \cite{bs2,sb}.

Let us consider the algebra of polynomials in $p$ with the finite highest
order:
\begin{equation}\label{dalg}
  \Alg = \Alg_{\me k-1} \oplus \Alg_{< k-1} = \pobr{\sum^N_{i\me k-1}
u_i(x)p^i} \oplus \pobr{\sum_{i< k-1} u_i(x)p^i}
\end{equation}
and with the Lie structure induced by the Poisson bracket in the form
\begin{equation}\label{pb}
  \pobr{f,g} := p\X(x)\bra{\pd{f}{p}\pd{g}{x}-\pd{f}{x}\pd{g}{p}}\qquad f,g\in\Alg.
\end{equation}
The subsets $\Alg_{\me k-1}$ and $\Alg_{< k-1}$ of $\Alg$ are Lie
subalgebras only for $k=1$ and $k=2$. Thus, the classical
$R$-matrices $R = P_{\me k-1} - \frac{1}{2}$ determine the Lax
hierarchies:
\begin{equation}\label{dlaxh}
L_{t_n} = \pobr{\bra{L^n}_{\me k-1},L}= \pi_0 dH_n =
\pi_1 dH_{n-1}\qquad n\in \Z_+\qquad k=1,2,
\end{equation}
that are generated by powers of the Lax functions $L\in \Alg$
given in the form:
\begin{align}
\label{dl1} k=1:\qquad L &= p+u_0+u_1p^{-1}+u_2p^{-2}+...=
p+\sum_{i\me 0}u_{i}p^{-i}\\
\label{dl2} k=2:\qquad L &= u_0p+u_1+u_2p^{-1}+u_3p^{-2}+...=
\sum_{i\me 0}u_{i}p^{1-i}.
\end{align}

The first dispersionless chains from \eqref{dlaxh} for $k=1$ take
the following form
\begin{align}\label{e1}
(u_i)_{t_1} &= \X\brac{(u_{i+1})_x+iu_i(u_0)_x}\\\notag
(u_i)_{t_2} &=
2\X\brac{(u_{i+2})_x+u_0(u_{i+1})_x+(i+1)u_{i+1}(u_0)_x+i
u_iu_0(u_0)_x+i u_i(u_1)_x}\\\notag
        & \vdots\ ,
\end{align}
and for $k=2$
\begin{align}\label{e2}
(u_i)_{t_1} &= \X\brac{u_0(u_{i+1})_x+iu_{i+1}(u_0)_x}\\\notag
(u_i)_{t_2} &=
2\X\brac{u_0^2(u_{i+2})_x+(i+1)u_0u_{i+2}(u_0)_x+u_0u_1(u_{i+1})_x
+iu_{i+1}(u_0u_1)_x}\\\notag
        & \vdots\ .
\end{align}

\begin{example}
For $\X=1$ the chains \eqref{e1} and \eqref{e2} are dispersionless
Toda and modified Toda chains, respectively, while for
$\X=x$ the chains \eqref{e1} and \eqref{e2} are
dispersionless limits of the $q$-analogues of Toda and modified
Toda.
\end{example}

The appropriate trace form is defined as
\begin{equation*}
  \tr(A) :=  \int_{-\infty}^{\infty} \X^{-1}\res (A p^{-1})\,dx,
\end{equation*}
where $\res (A):=a_{-1}$ for $A=\sum_i a_i p^i$, and the inner
product on $\Alg$ is given by
\begin{equation*}
(A,B):= \tr(A B).
\end{equation*}
\begin{proposition}
The above inner product is nondegenerate, symmetric and
$\ad$-invariant with respect to the Poisson bracket, i.e.
\begin{equation*}
  \bra{\pobr{A,B},C} = \bra{A,\pobr{B,C}}\qquad A,B,C\in\Alg.
\end{equation*}
\end{proposition}
\begin{proof}
The nondegeneracy and symmetricity is obvious. The
$\ad$-invariance is a consequence of the following equality: $\tr
\pobr{A,B} = 0$, which is valid for arbitrary $A,B\in\Alg$.
\end{proof}

Then, the differentials $dH(L)$ of functionals $H(L)\in \F(\Alg)$
related to the Lax functions \eqreff{dl1}{dl2} have the form:
\begin{align*}
  k=1:\qquad dH&= \X\sum_{i\me 0}\var{H}{u_i}p^i\\
  k=2:\qquad dH&= \X\sum_{i\me 0}\var{H}{u_i}p^{i-1}.
\end{align*}

The bi-Hamiltonian structure of the Lax hierarchies \eqref{laxh}
is defined through the compatible (for fixed $k$) Poisson tensors:
\begin{align*}
k=1,2:\qquad\pi_0: dH\map
\pobr{L,(dH)_{<k-1}}+\bra{\pobr{dH,L}}_{< 2-k}
\end{align*}
and
\begin{align*}
&k=1:\qquad\pi_1: dH\map \pobr{L,\bra{dH L}_{<0}}
+ L\bra{\pobr{dH,L}}_{<1}+\pobr{\pr_x^{-1}\res\bra{\X^{-1}\pobr{dH,L}p^{-1}},L}\\
&k=2:\qquad\pi_1: dH\map \pobr{L,\bra{dH L}_{<1}} + L\bra{\pobr{dH,L}}_{<0}.
\end{align*}
Then, for Hamiltonians
\begin{equation*}
  H_n(L) = \frac{1}{n+1}\tr\bra{L^{n+1}}\qquad dH_n(L)=L^n,
\end{equation*}
the explicit bi-Hamiltonian structure of \eqref{laxh} is
given by
\begin{equation*}
  (u_i)_{t_n} = \sum_{j\me 0}\pi_0^{ij}\,\var{H_n}{u_j}
  = \sum_{j\me 0}\pi_1^{ij}\,\var{H_{n-1}}{u_j}\qquad i\me 0.
\end{equation*}
So, the Poisson tensors for $k=1$ are given by
\begin{align*}
\pi_0^{ij} &= \X\brac{j\pr_xu_{i+j}+iu_{i+j}\pr_x}\X\\
\pi_1^{ij} &=
\X\Big[\sum_{k=0}^i\bra{(j-k)u_k\pr_xu_{i+j-k}+(i-k)u_{i+j-k}\pr_xu_k}
+ i(j+1)u_i\pr_xu_j\\
&\qquad + (j+1)\pr_xu_{i+j+1}+(i+1)u_{i+j+1}\pr_x\Big]\X
\end{align*}
where the related Hamiltonians are
\begin{align*}
    H_0 &= \int_{-\infty}^\infty \X^{-1}u_0\,dx\\
    H_1 &= \int_{-\infty}^\infty \X^{-1}\bra{u_1+\frac{1}{2}u_0^2} dx\\
    H_2 &= \int_{-\infty}^\infty
    \X^{-1}\bra{u_2+2u_0u_1+\frac{1}{3}u_0^3} dx\\
    &\vdots\ .
\end{align*}
For $k=2$ we have the first Poisson tensor
\begin{align*}
&\pi_0^{10} = \X\pr_x\X u_0\qquad \pi_0^{01} = u_0\X\pr_x\X\\
&\pi_0^{ij} = \X\brac{(j-1)\pr_xu_{i+j-1}+(i-1)u_{i+j-1}\pr_x}\X\quad i,j\me
2,
\end{align*}
where all remaining $\pi_0^{ij}$ are equal zero, and the second one
\begin{align*}
\pi_1^{ij} = \X\Big[\sum_{k=0}^{i-1}
\bra{(j-k)u_k\pr_xu_{i+j-k}+(i-k)u_{i+j-k}\pr_xu_k} +
(1-i)u_i\pr_xu_j\Big]\X.
\end{align*}
Finally
\begin{align*}
  H_0 &= \int_{-\infty}^\infty \X^{-1}u_1\,dx\\
  H_1 &= \int_{-\infty}^\infty \X^{-1}\bra{\frac{1}{2}u_1^2+u_0u_2} dx\\
  H_2 &= \int_{-\infty}^\infty \X^{-1}\bra{\frac{1}{3}u_1^3+u_0^2u_3+2u_0u_1u_2} dx\\
  &\vdots\ .
\end{align*}

One can observe that the chains, together with the bi-Hamiltonian
structures, constructed in this section are dispersionless limits
of the discrete chains considered in Section 3.

\section{Deformation quantization procedure}

The aim of this section is formulation of the inverse procedure to
the dispersionless limit considered earlier. Using the formalism
of quantization deformation (for the references see \cite{bs}) the
unified approach to the lattice and field soliton systems was
presented in \cite{bs}. Here we follow the scheme from that
article.

The Poisson bracket \eqref{pb} can be written in the form
\begin{equation*}
  \pobr{f,g} := f\bra{p\pr_p\wedge\X(x)\pr_x}g\qquad f,g\in\Alg,
\end{equation*}
where the derivations $p\pr_p$ and $\X(x)\pr_x$ commute. Hence, it
can be quantized in infinitely many ways via $\star$-products
being deformed multiplications
\begin{equation}\label{star}
  f\star^\alpha g = f\exp\brac{\frac{\hk}{2}\bra{(\alpha+1)p\pr_p\otimes\X(x)\pr_x
  +(\alpha-1)\X(x)\pr_x\otimes p\pr_p}}.
\end{equation}
This $\star$-product for $\alpha=0$ and $\alpha=1$ is the
generalization of the Moyal and Kuperschmidt-Manin products,
respectively. Expanding \eqref{star} one finds that
\begin{align}\label{starexp}
  f\star^{\alpha}g =
  \sum_{k=0}^\infty\frac{\hk^k}{2^kk!}\sum_{j=0}^k(\alpha+1)^{k-j}(\alpha-1)^j
  \brac{(p\pr_p)^{k-j}(\X\pr_x)^jf}\cdot\brac{(\X\pr_x)^{k-j}(p\pr_p)^jg}.
\end{align}

The algebra $\Alg$ \eqref{dalg} with the multiplication defined as
\eqref{star}, with fixed $\alpha$, is the associative, but not
commutative, algebra with Lie bracket, being a deformed Poisson
bracket, defined as
\begin{equation}\label{deflie}
  \pobr{f,g}_{\star^\alpha} = \frac{1}{\hk}\bra{f\star^\alpha g-g\star^\alpha
  f}.
\end{equation}
Then, in the limit $\hk\arrow 0$, we have that
\begin{align*}
  &\lim_{\hk\arrow 0}f\star^\alpha g = fg\\
  &\lim_{\hk\arrow 0}\pobr{f,g}_{\star^\alpha} = \pobr{f,g}.
\end{align*}
The algebra $\Alg$ with $\star^\alpha$-product will be denoted as
$\Alg_\alpha$.

The associativity property of $\star^\alpha$-products is a purely
algebraic consequence of their construction. For the simple proof
see \cite{bs}. Moreover, we could treat these products only
formally not requiring a convergence of the sum in
\eqref{starexp}. In order to make the $\star^\alpha$-products
consistent with the introduced formalism of grain structures we
assume that vector fields $\X\pr_x$ are such that the formula
\eqref{fact1} is valid. From the simple observation:
\begin{equation*}
(p\pr_p)^k p^m = m^kp^m,
\end{equation*}
one finds that
\begin{align*}
  p^m\star^\alpha u(x) &=
  \sum_{k=0}^\infty\frac{\hk^k}{2^kk!}(\alpha+1)^km^k
  (\X\pr_x)^ku(x)\, p^m = e^{m(\alpha+1)\frac{\hk}{2}\X\pr_x}u(x)\, p^m
  = E^{m\frac{\alpha+1}{2}}u(x)\, p^m\\
u(x)\star^\alpha p^m &=
  \sum_{k=0}^\infty\frac{\hk^k}{2^kk!}(\alpha-1)^km^k
  (\X\pr_x)^ku(x)\, p^m = e^{m(\alpha-1)\frac{\hk}{2}\X\pr_x}u(x)\, p^m
  = E^{m\frac{\alpha-1}{2}}u(x)\, p^m,
\end{align*}
where the last equalities follow from the above assumption and
\eqref{fact}.

It is important that the decomposition of \eqref{dalg} into Lie
subalgebras after deformation quantization is preserved and they
are still Lie subalgebras with respect to the Lie bracket
\eqref{deflie}. Hence, we have Lax hierarchies
\begin{equation}\label{deflaxh}
L_{t_n} = \pobr{\bra{L^n}_{\me k-1},L}_{\star^\alpha}\qquad n\in
\Z_+\qquad k=1,2,
\end{equation}
which be still well-defined for Lax functions in the form
\eqreff{dl1}{dl2}. Notice that now, the Lax hierarchies are
generated by powers with respect to $\star^\alpha$-products, i.e.
$L^n = L\star^\alpha...\star^\alpha L$. The first chains from Lax
hierarchies \eqref{deflaxh} are
\begin{align*}
  k=1:\qquad (u_i)_{t_1} &= \frac{1}{\hk}\brac{(E-1)E^{\frac{\alpha-1}{2}}u_{i+1}
  +u_i(1-E^{-i})E^{i\frac{1-\alpha}{2}}u_0}\\
  k=2:\qquad (u_i)_{t_1} &= \frac{1}{\hk}\brac{E^{i\frac{1-\alpha}{2}}u_0E^{\frac{\alpha+1}{2}}u_{i+1}
  -E^{\frac{\alpha-1}{2}}u_{i+1}E^{-i(\frac{\alpha+1}{2})}u_0}.
\end{align*}
One can observe that they coincide with the respective discrete
systems from Section \ref{sect} for $\alpha=1$.

Nevertheless, all algebras $\Alg_\alpha$ are gauge equivalent
under the isomorphism
\begin{equation*}
  D^{\alpha'-\alpha}:\Alg_\alpha\arrow \Alg_{\alpha'}\qquad
  D^{\alpha'-\alpha}=
  \exp\brac{(\alpha-\alpha')\frac{\hk}{2}\,\X(x)\pr_x\,p\pr_p},
\end{equation*}
such that
\begin{align*}
f\star^{\alpha'} g &=
D^{\alpha'-\alpha}\brac{D^{\alpha-\alpha'}f\star^\alpha
D^{\alpha-\alpha'}g}\\
\pobr{f,g}_{\star^{\alpha'}} &=
D^{\alpha'-\alpha}\pobr{D^{\alpha-\alpha'}f,D^{\alpha-\alpha'}g}_{\star^\alpha}.
\end{align*}
It is also straightforward to prove that under the above
isomorphism the Lax hierarchy structure is preserved. Let
$L_\alpha = \sum_i u_ip^i\in \Alg_\alpha$ and $L_{\alpha'} =
\sum_i u_i'p^i\in \Alg_{\alpha'}$. Then, the transformation
between fields follows
\begin{equation*}
  L_{\alpha'} = D^{\alpha'-\alpha}L_\alpha\qquad\Rightarrow\qquad
  u_i'= E^{i\frac{\alpha-\alpha'}{2}}u_i.
\end{equation*}

On the other hand, directly from \eqref{star} the following
commutation rules result:
\begin{align*}
  &u\star v = uv\\
  &p^m\star p^n = p^{m+n}\\
  &p^m\star u = \bra{e^{m\hk\X\pr_x}u}\star p^m = E^mu\star p^m\\
  &u\star p^m = p^m\star\bra{e^{-m\hk\X\pr_x}u} = p^m\star E^{-m}u,
\end{align*}
being independent of the choice of $\star^\alpha$-product,
therefore we have skipped the related index. Hence, we can
quantize separately the algebra $\Alg$ to the following algebra
\cite{dp}
\begin{equation*}
  \alga = \pobr{\sum_i u_i\star p^i},
\end{equation*}
which obviously is associative under the above commutation rules.
Notice that the algebra $\alga$ differs from algebras
$\alg_\alpha$ as in $\alga$ we also deformed the polynomial
functions, i.e. we are not using the standard multiplication any
more. Let us point out that the algebra $\alga$ is trivially
equivalent to the algebra $\Alg_1$ as $u\star^1 p^m = u p^m$ and
$p^m\star^1 u = E^mu p^m$. Also, it is straightforward to see that
$\alga$ is isomorphic to the algebra of shift operators $\alg$
\eqref{alg} defined on the grain structure by some discrete
one-parameter group of diffeomorphisms on $\Rm$. Hence, it is
clear that the algebra \eqref{dalg} with Poisson bracket
\eqref{pb} is the limit, $\hk\arrow 0$, of the algebra \eqref{alg}
of shift operators with the Lie structure defined by the
commutator.

\section{Conclusions}

In the present article we have introduced a general framework of
integrable discrete systems on $\Rm$. This formalism is based on
the construction of shift operators by means of discrete
one-parameter groups of diffeomorphisms on $\Rm$ that are
determined by infinitesimal generators $\X\pr_x$. Particularly, if
$\X=1$ or $\X=x$ the discrete systems considered are of lattice
Toda or $q$-deformed Toda type, respectively. Although the
construction of integrable systems related to different $\X\pr_x$
is completely parallel, they, in general, are not equivalent,
similarly as the vector fields on $\Rm$ are not equivalent.

The two vector fields $\X(x)\pr_x$, $\X'(x')\pr_{x'}$ and the
related discrete systems are equivalent if integrating
\begin{equation*}
  \int\frac{dx}{\X(x)} = \int\frac{dx'}{\X'(x')}
\end{equation*}
one finds a bijective map between $x$ and $x'$, otherwise related
one-parameter groups of diffeomorphisms do not transform into one
another.

Consider the vector fields from Example~\ref{example}. Let $\X(x)
= x^{1-n}$ for $n\neq 0$ and $\X'(x')=1$ (the lattice case). Then
one finds that $x'=\frac{1}{n}x^n$, which is a bijection for $n$
being odd. Hence, all the discrete systems generated by
$\X\pr_x=x^{1-n}\pr_x$ with odd $n$ can be reduced to the original
lattice Toda type systems. For $n=0$ $\X(x)=x$ (the $q$-discrete
case) and let again $\X'(x')=1$. Then $x=e^{x'}$ is not the
bijection. However, if the domain of dynamical fields of
$q$-discrete systems is restricted to $x\in \Rm_+$, then the above
map is a bijection and $q$-discrete systems on $\Rm_+$ became
equivalent to the lattice systems on $\Rm$.

\subsection*{Acknowledgement}

The work of M. B\l aszak, M. G\"urses and B. Silindir was
partially supported by the Turkish Academy of Sciences and by the
Scientific and Technological Research Council of Turkey. Moreover
M. B\l aszak and B. Szablikowski were supported by MNiI research
grant no. N202.


\footnotesize
\begin{thebibliography}{99}

\bibitem{k} C. Kassel, {\it Cyclic homology of differential
operators, the Virasoro algebra and a $q$-analogue}, Comm. Math.
Phys. {\bf 146} (1992) 343-356

\bibitem{wzz} D.H. Zhang, {\it Quantum
deformation of KdV hierarchies and their infinitely many
conservation laws}, J. Phys. A: Math. Gen. {\bf 26} (1993)
2389-2407

\bibitem{ms} J. Mas and M. Seco, {\it The algebra of
$q$-pseudodifferential symbols and the $q-W^{(n)}_{\rm KP}$
algebra}, J. Math. Phys. {\bf 37} (1996) 6510-6529

\bibitem{f} E. Frenkel, {\it Deformations of the KdV hierarchy
and related soliton equations}, Int. Math. Res. Not. {\bf 2}
(1996) 55-76

\bibitem{klr} B. Khesin, V. Lyubashenko and C. Roger, {\it
Extensions and contractions of the Lie algebra of
$q$-pseudodifferential symbols on the circle}, J. Funct Anal. {\bf
143} (1997) 55-97

\bibitem{ahm} M. Adler, E. Horozov and P. van Moerbeke, {\it The
solution to the $q$-KdV equation}, Phys. Lett. {\bf A 242} (1998)
139-151

\bibitem{t} K. Takasaki, {\it $q$-Analogue of Modified KP
Hierarchy and its Quasi-Classical Limit}, Lett. Math. Phys. {\bf
70} (2005) 165-181

\bibitem{hlc} J. He, Y. Li and Y. Cheng, {\it $q$-Deformed KP
Hierarchy and $q$-Deformed Constrained KP Hierarchy}, SIGMA {\bf
2} (2006) Paper 060

\bibitem{dmh} A. Dimakis and F. M\"uller-Hoissen, {\it Functional
representations of integrable hierarchies}, arXiv:nlin.SI/0603018


\bibitem{sts} M. A. Semenov-Tian-Shansky, {\it What is a classical
r-matrix?}, Funct. Anal. Appl. {\bf 17} (1983) 259

\bibitem{ko} B. G. Konopelchenko and W. Oevel, {\it An r-matrix approach
to nonstandard classes of integrable equations}, Publ. RIMS, Kyoto
Univ. {\bf 29} (1993) 581-666

\bibitem{bm} M. B\l aszak and K. Marciniak, {\it R-matrix approach to lattice integrable
systems}, J. Math. Phys. {\bf 35} (1994) 4661

\bibitem{o} W. Oevel,{\it Poisson Brackets in Integrable Lattice Systems} in
{\it Algebraic Aspects of Integrable Systems} edited by A.S. Fokas
and I.M. Gelfand, Progress in Nonlinear Differential Equations Vol.
26 (Birkh\"{a}user-Boston) (1996) 261

\bibitem{b} M. B\l aszak, {\it Multi-Hamiltonian Theory of Dynamical Systems},
Texts and Monographs in Physics (Springer-Verlag, Berlin, 1998) 350 pp.

\bibitem{bs} M. B\l aszak and B. M. Szablikowski, {\it From dispersionless to soliton
systems via Weyl-Moyal-like deformations}, J. Phys A: Math. Gen.
{\bf 36} (2003) 12181-12203

\bibitem{tt} K. Takasaki and T. Takebe, {\it Integrable
hierarchies and dispersionless limit}, Rev. Math. Phys. {\bf 7}
(1995) 743-808

\bibitem{li} Li Luen-Chau, {\it Classical r-Matrices and Compatible Poisson
Structures for Lax Equations in Poisson Algebras}, Commun. Math.
Phys. {\bf 203} (1999) 573-592

\bibitem{bs2} M. B\l aszak and B. M. Szablikowski, {\it Classical $R$-matrix theory
of dispersionless systems: I. (1+1)-dimension theory}, J. Phys A:
Math. Gen. {\bf 35} (2002) 10325-10344

\bibitem{sb} B. M. Szablikowski and M. B\l aszak, {\it Meromorphic Lax representations
of (1+1)-dimensional multi-Hamiltonian dispersionless systems}, J.
Math. Phys. {\bf 47} (2006) 092701

\bibitem{ah} B. Aulbach, S. Hilger,{\it Linear dynamic processes with
inhomogeneous time scale}, Nonlinear dynamics and quantum dynamical
systems (Gaussig, 1990) Math. Res. 59 (Akademie-Verlag, Berlin,
1990), pp. 9-20

\bibitem{ggs} M. G\"urses, G. Sh. Guseinov and B. Silindir, {\it Integrable
equations on time scales}, J. Math. Phys {\bf 46} (2005) 113510

\bibitem{cdz} G. Carlet, B. Dubrovin and Y. Zhang, {\it The
Extended Toda Hierarchy}, Mosc. Math. J. {\bf 4} (2004) 313-332

\bibitem{dp} A. Das and Z. Popowicz, {\it Properties of Moyal-Lax
representation}, Phys. Lett. B  {\bf 510}  (2001) 264-270

\end{thebibliography}
\end{document}